\documentclass{elsart}
\usepackage[square,sort&compress,comma]{natbib}
\usepackage{color,setspace,times}
\usepackage{amssymb,natbib,amsmath,graphicx,color,rotating,subfigure,url}
\usepackage{lineno}

\journal{Physica A}

\begin{document}

\begin{frontmatter}

\title{Scaling and Memory Effect in Volatility
Return Interval of the Chinese Stock Market}
\author[SE]{T. Qiu \corauthref{cor}},
\corauth[cor]{Corresponding author. Address: 696 South Fenghe
Avenue, School of Electronic and Information Engineering, Nanchang
Hangkong University, Nanchang, 330063, China.}
\ead{tianqiu.edu@gmail.com} %
\author[SB]{L. Guo},
\author[SE]{G. Chen}

\address[SE]{School of Electronic and Information Engineering, NanChang Hangkong University, Nanchang, 330063, China}
\address[SB]{School of Business, East China University of Science and Technology, Shanghai, 200237, China}

\begin{abstract}
Abstract: We investigate the probability distribution of the
volatility return intervals $\tau$ for the Chinese stock market. We
rescale both the probability distribution $P_{q}(\tau)$ and the
volatility return intervals $\tau$ as $P_{q}(\tau)=1/\overline{\tau}
f(\tau/\overline{\tau})$ to obtain a uniform scaling curve for
different threshold value $q$. The scaling curve can be well fitted
by the stretched exponential function $f(x) \sim e^{-\alpha
x^{\gamma}}$, which suggests memory exists in $\tau$. To demonstrate
the memory effect, we investigate the conditional probability
distribution $P_{q} (\tau|\tau_{0})$, the mean conditional interval
$\langle\tau|\tau_{0}\rangle$ and the cumulative probability
distribution of the cluster size of $\tau$. The results show clear
clustering effect. We further investigate the persistence
probability distribution $P_{\pm}(t)$ and find that $P_{-}(t)$
decays by a power law with the exponent far different from the value
0.5 for the random walk, which further confirms long memory exists
in $\tau$. The scaling and long memory effect of $\tau$ for the
Chinese stock market are similar to those obtained from the United
States and the Japanese financial markets.

\end{abstract}


\end{frontmatter}

\section{Introduction}

%\hspace{3mm} New paragraph

In recent years, physicists have paid much attention on the dynamics
of financial markets. Scaling behavior is discovered in the
financial system by analyzing the indices and the stock prices
$y(t')$, such as the 'fat tail' of the probability distribution
$P(Z,t)$of the two point price change return $Z(t')=ln\ y(t')-ln\
y(t'-1)$\cite {man95,gop99}. The physical origin of the scaling
behavior is often related to the long range correlation. It is
interested to find, in spite of the absence of the return
correlation, the volatility $|Z(t')|$ is long range correlated \cite
{gia01,liu99}.

Recently, the volatility return intervals $\tau$, which is defined
as the return intervals that the volatility is above a certain
threshold $q$, is investigated for the United States and the
Japanese financial markets \cite
{yam05,wan06,wan07,jun08,vod08,wan08}. Scaling behavior of the
probability distribution in the volatility return intervals $\tau$
is discovered, and long-range autocorrelation is demonstrated for
$\tau$. The scaling and the long-range autocorrelation are rather
robust independent of the stock markets and the foreign exchange
markets for the developed countries. However, it is known that the
emerging markets may behave differently \cite {qiu06,qiu07,zho04}.
Especially, the Chinese stock market is newly set up in 1990 and
shares a transiting social and political system. Due to the special
background of the Chinese stock market, it may share similar
properties as the mature financial markets \cite {qiu07}, however,
it may also exhibits special features far different from the mature
financial markets in some aspects \cite {qiu06,qiu07,zho04}, such as
the leverage effect reported in ref. \cite {qiu06,qiu07}. It is
important to investigate the financial dynamics for the Chinese
stock market to achieve more comprehensive understanding of the
financial markets.

In this paper, to broaden the understanding of the scaling and
memory effect of the volatility return intervals $\tau$ for the
emerging markets, we investigate the probability distribution and
the memory effect of $\tau$ for the Chinese stock market. In the
next section, we present the data set we analyzed, In section 3, we
show the probability distribution of $\tau$. In section 4, we
investigate the clustering phenomena by analyzing the conditional
probability distribution $P_{q} (\tau|\tau_{0})$, the mean
conditional interval $\langle\tau|\tau_{0}\rangle$ and the
cumulative distribution of the cluster size of $\tau$. In section 5,
we investigate the persistence probability distribution
$P_{\pm}(t)$. Finally comes the conclusion.

\section{Data Analyzed}
The data we analyzed is based on the trade-by-trade data from the
stocks of the Shanghai Stock Exchange market(SHSE) and the Shenzhen
Stock Exchange market(SZSE). The SHSE was established on November
26, 1990 and put into operation on December 19, 1990. Shortly after,
the SZSE was established on December 1, 1990 and put into operation
on July 3, 1991. Most A-shares and B-shares are traded in the SHSE
and SZSE.

The Chinese stock market is an order-driven market and is based on
the so called continuous double auction mechanism. In the trading
day, there are 3 time periods. From 9:15 to 9:25 a.m., it is the
opening call auction time, when the buy and sell orders are
aggregated to match. From 9:25 to 9:30 a.m., it is the cool period,
and then followed by the continuous double auction time. The time
period for the continuous double auction is from 9:30 to 11:30 a.m.
and from 13:00 to 15:00 p.m.. More Information about the development
process and the current trading mechanism can be found in ref. \cite
{zho04,gu07,gu08a,gu08b}. We study the transaction records for 3
whole year from 2003 to 2006. The number of the 3 whole year
transactions is about 800,000 on average.

\section{The Probability Distribution}
Here we define the volatility return intervals $\tau(q)$ as the time
intervals that volatility $|Z(t')|$ above a certain threshold $q$,
where the sampling time interval for the volatility $|Z(t')|$ is 1
min. Therefore, $\tau(q)$ depends on the threshold $q$. Fig. 1 shows
the volatility return intervals for $q=0.50$, $q=1.00$ and $q=1.50$
in May 2003 of the Datang Telecom Co., Ltd(DTT). The big value of
$q$ corresponds to the large volatility that rarely occurs in the
financial markets. We investigate the probability distribution
function(PDF) $P_{q}(\tau)$ of the volatility return interval
$\tau(q)$ with the threshold $q=0.750$, 0.875, 1.000, 1.125, 1.375
and 1.500.

\begin{figure}[htb]
\centering
\includegraphics[width=8.5cm]{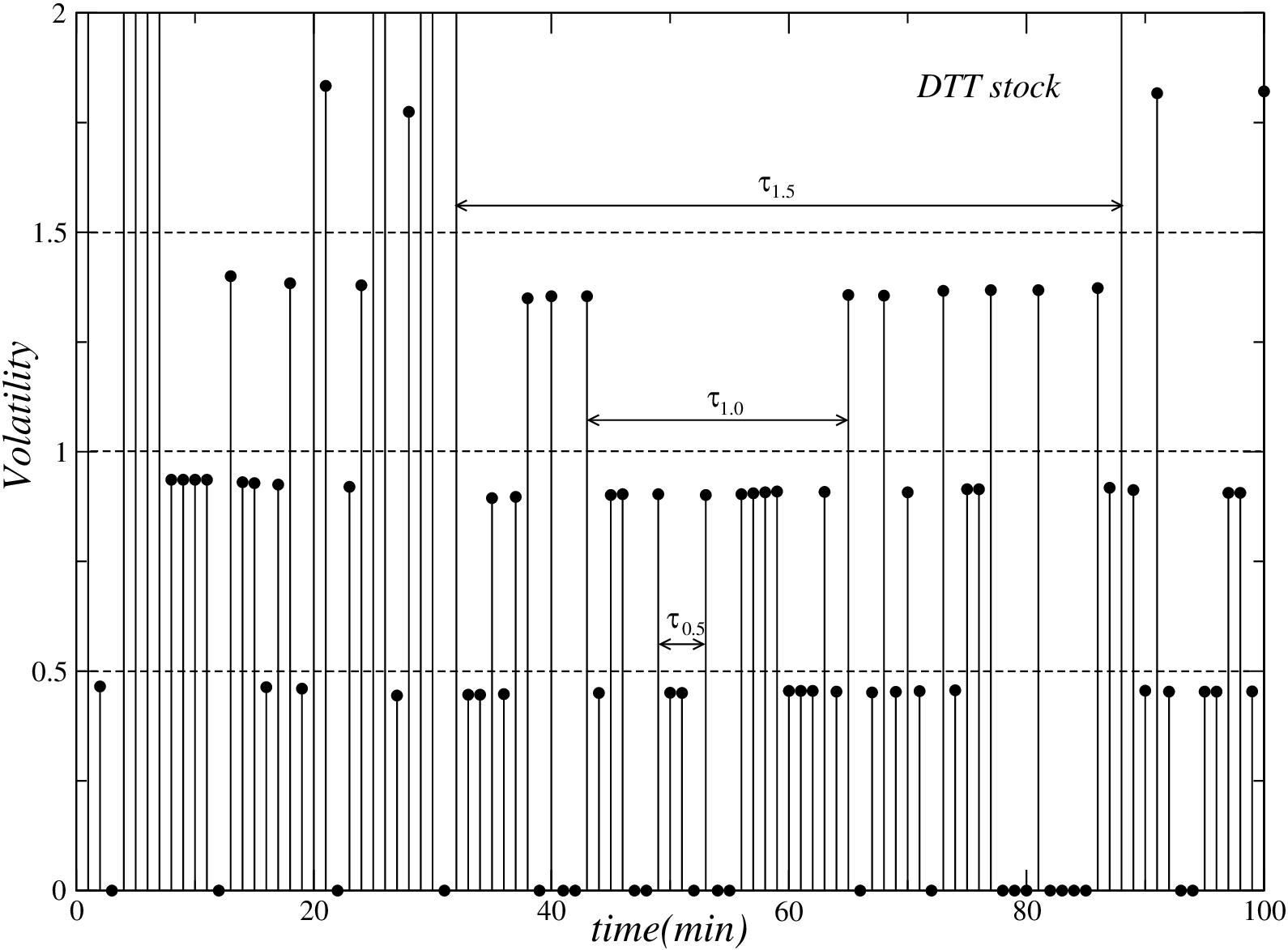}
\caption{\label{Fig:1} Volatility return intervals $\tau$ for the
DTT stock in May 2003 with the threshold values $q$ =0.50, 1.00,
1.50 are displayed.}
\end{figure}

Fig. 2a shows the PDF $P_{q}(\tau)$ for the DTT stock and Fig. 2b
shows the PDF $P_{q}(\tau)$ for the Chinese Minsheng Banking Co.,
Ltd(CMB). The seven curves are for $q$=0.750, 0.875, 1.000, 1.125,
1.375 and 1.500 respectively. The results show that the PDF
$P_{q}(\tau)$ for large $q$ decays slower than that for small $q$.
We rescale $P_{q}(\tau)$ and $\tau$ as
$P_{q}(\tau)=1/\overline{\tau} f(\tau/\overline{\tau})$, which is
mentioned in ref. \cite {yam05,wan06,wan07,jun08,vod08,wan08}, to
collapse the seven curves with different threshold $q$ onto a single
curve, where $\overline{\tau}$ is the average interval.

\begin{figure}[htb]
\centering
\includegraphics[width=8.5cm]{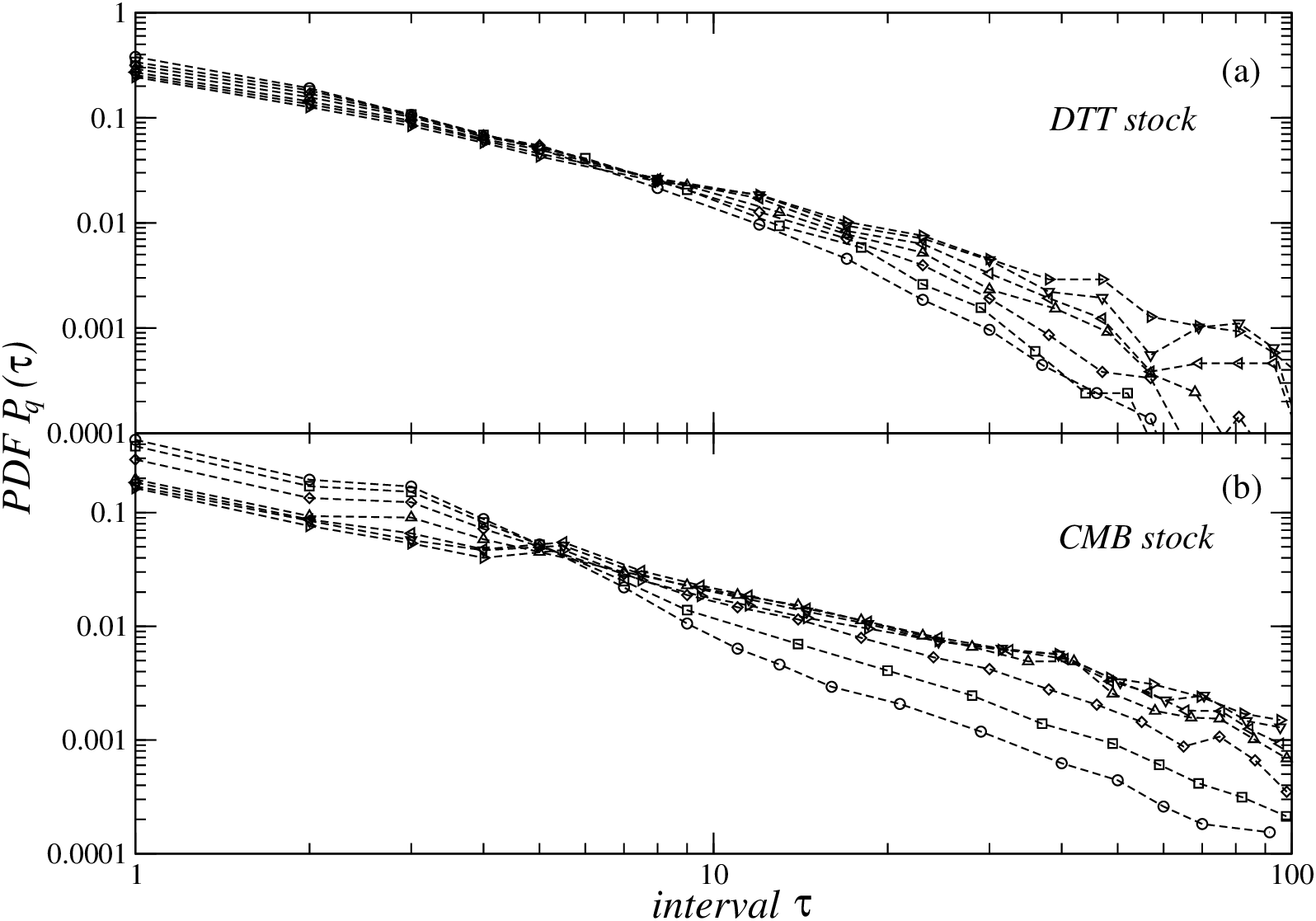}
\caption{\label{Fig:2} (a) Probability distribution functions
$P_{q}(\tau)$ for the DTT stock are displayed in log-log scale. The
circles, squares, diamonds, triangle ups, triangle lefts, triangle
downs and triangle rights are for the threshold values $q$ from 0.75
to 1.50 (0.750, 0.875, 1.000, 1.125, 1.250, 1.375 and 1.500)
respectively. (b) Probability distribution functions $P_{q}(\tau)$
for the CMB stock are displayed in log-log scale. The  circles,
squares, diamonds, triangle ups, triangle lefts, triangle downs and
triangle rights are for the threshold values $q$ from 0.75 to 1.50
(0.750, 0.875, 1.000, 1.125, 1.250, 1.375 and 1.500 respectively.}
\end{figure}

Fig. 3a and Fig. 3b shows the scaled PDF
$P_{q}(\tau)\overline{\tau}$ as a function of the scaled volatility
return intervals $\tau/\overline{\tau}$ for the DTT stock and the
CMB stock. Fig. 3c shows the scaled PDF $P_{q}(\tau)\overline{\tau}$
for the DTT stock, the CMB stock, the CITIC Securities Co., Ltd.
(CITIC) stock and the Bird Telecom  Co., Ltd(BDT) stock.

\begin{figure}[htb]
\centering
\includegraphics[width=8.5cm]{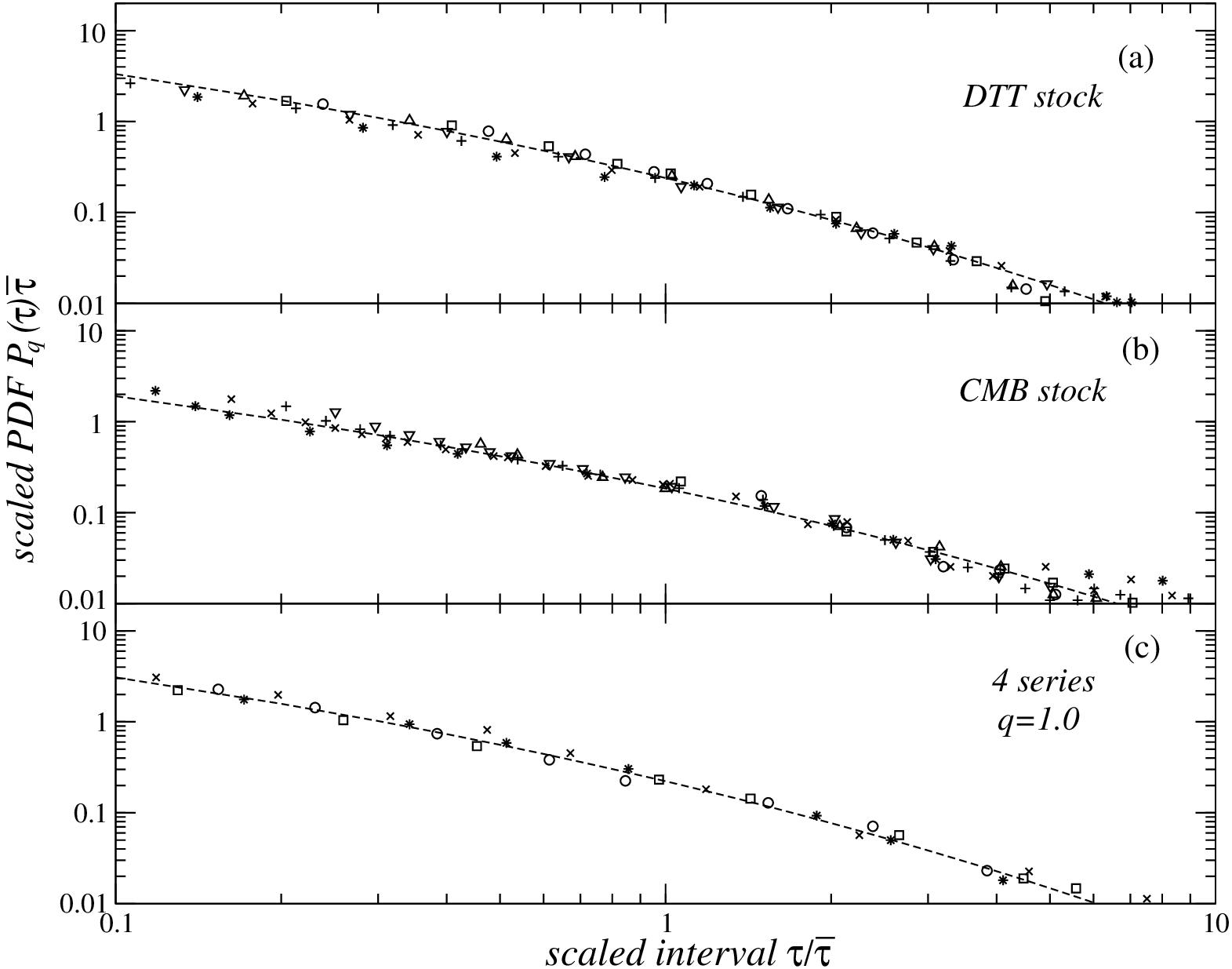}
\caption{\label{Fig:3} (a) Scaling of the volatility return
intervals for the DTT stock with the threshold values $q$ ranging
from 0.75 to 1.50 (0.750, 0.875, 1.000, 1.125, 1.250, 1.375 and
1.500) are displayed with circles, squares, triangle ups, triangle
downs, pluses, crosses and stars in log-log scale. (b) Scaling of
the volatility return intervals for the CMB stock with the threshold
values $q$ ranging from 0.75 to 1.50 (0.750, 0.875, 1.000, 1.125,
1.250, 1.375 and 1.500) are displayed with circles, squares,
triangle ups, triangle downs, pluses, crosses and stars in log-log
scale. (c) Scaling of the volatility return intervals for the DTT
stock, the CMB stock, the CITIC stock and the BDT stock with the
threshold value $q=1.0$ are displayed with circles, squares, crosses
and stars in log-log scale.}
\end{figure}

The scaling behavior is observed for the normalized $\tau$. The
scaling function $f(\tau/\overline{\tau})$ does not directly depend
on the threshold $q$ but through
$\overline{\tau}\equiv\overline{\tau}(q)$. The scaling behavior is
similar to that obtained from the United States and the Japanese
stock markets, i.e., the generality of the scaling is further
confirmed for both the mature financial markets and the Chinese
stock market. It helps us to overcome the difficulty to perform
statistics for the rare event with big price fluctuation. If we know
the $P_{q}(\tau)$ with a small $q$, the behavior of $\tau$ with
large $q$ then can be predicted by the scaling function. We fit the
scaled PDF with a stretched exponential function form \cite{bun05},

\begin{equation}
f(x) \sim e^{-\alpha x^{\gamma}},
\end{equation}

We find that the DTT stock and the CMB stock have similar exponent
value with $(\gamma, \alpha)=(0.20\pm0.05, 4.0\pm0.5)$. The exponent
is close to the value $(\gamma, \alpha)=(0.38\pm0.05, 3.9\pm0.5)$ of
the United States stock market \cite {wan06}. The function form is
far different from the Poission distribution, which indicates there
may exist correlation in $\tau$.

\section{Clustering Phenomena of the Volatility Return Intervals}
If clustering effect occurs in time serials, it suggests memory
exists in those serials. To demonstrate the memory in the volatility
return intervals $\tau$, we investigate the clustering effect by
study the conditional probability distribution $P_{q}
(\tau|\tau_{0})$, the mean conditional interval
$\langle\tau|\tau_{0}\rangle$ and the cumulative probability
distribution of the cluster size of $\tau$.
\subsection{Conditional Probability Distribution}
To investigate the memory effect of the volatility return intervals
$\tau$, we analyze the conditional PDF $P_{q} (\tau|\tau_{0})$.  The
$P_{q} (\tau|\tau_{0})$ is denoted as the probability distribution
function of the $\tau$ that immediately follow a given volatility
return interval $\tau_{0}$
\cite{yam05,wan06,wan07,jun08,vod08,wan08}. If memory exists in
$\tau$, the $P_{q} (\tau|\tau_{0})$ should depend on the preceding
volatility return interval $\tau_{0}$. To achieve good statistics
with more data points, we sort the volatility return intervals in
increasing direction and divide it into two subsets. Fig. 4 shows
the scaled conditional PDF $P_{q} (\tau|\tau_{0})/\overline{\tau}$
for the DTT stock and the CMB stock with $q$=0.750, 0.875, 1.000,
1.125, 1.250, 1.375 and 1.500. The $P_{q}
(\tau|\tau_{0})/\overline{\tau}$ for different threshold $q$ are
collapsed onto a single curve. It is observed that the $P_{q}
(\tau|\tau_{0})/\overline{\tau}$ for the lower subset is higher for
the small $\tau_{0}$ while  the $P_{q}
(\tau|\tau_{0})/\overline{\tau}$ for the larger subset is higher for
the large $\tau_{0}$, i.e., small $\tau$ follows small $\tau_{0}$
and large $\tau$ follows large $\tau_{0}$. It implies that the
intervals with the similar size form clusters, i.e., there exists
memory in $\tau$. The clustering effect has been investigated in
ref. \cite {yam05,wan06,wan07,jun08,vod08,wan08}, we obtain the
similar result for the Chinese stock market. To guide the eye, we
fit the data with the stretched exponential function $f(x) \sim
e^{-\alpha x^{\gamma}}$. For the DTT stock, the exponent is measured
to be $(\gamma,\alpha)=(0.30,4.00)$ for the Lower 1/2 subset and
(0.20,4.00) for the larger 1/2 subset. For the CMB stock, the
exponent is measured to be $(\gamma,\alpha)=(0.30,4.0)$ for the
Lower 1/2 subset and (0.20,3.50) for the larger 1/2 subset.

\begin{figure}[htb]
\centering
\includegraphics[width=8.5cm]{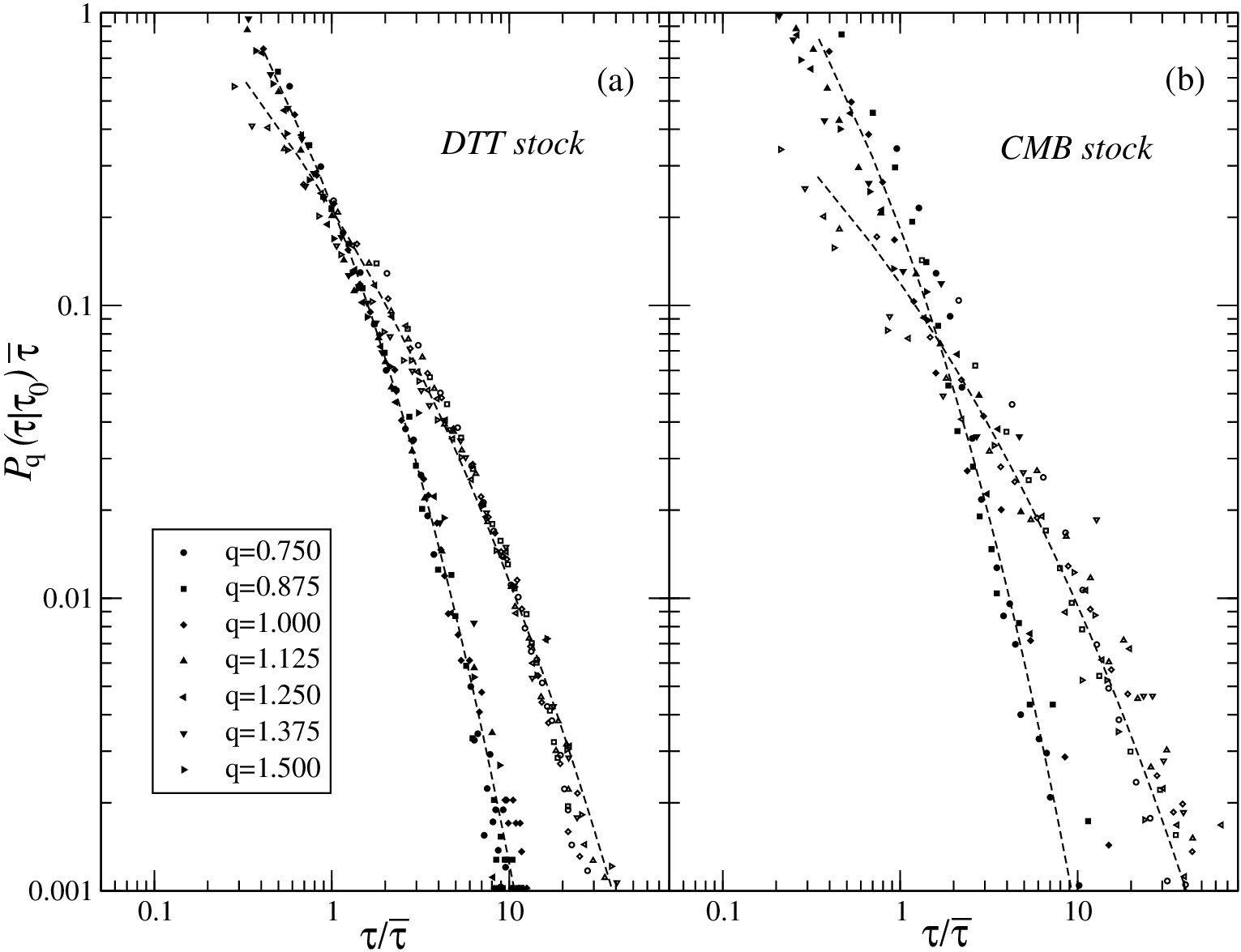}
\caption{\label{Fig:4} (a) The scaled conditional probability
distribution $P_{q} (\tau|\tau_{0})/\overline{\tau}$ vs
$\tau/\overline{\tau}$ for the DTT stock with the threshold values
$q$ ranging from 0.75 to 1.50 (0.750, 0.875, 1.000, 1.125, 1.250,
1.375 and 1.500) are displayed with circles, squares, diamonds,
triangle ups, triangle lefts, triangle downs and triangle rights.
The closed symbols are for the lower 1/2 subset, and  the open
symbols are for the larger 1/2 subset. The dashed lines are for
guiding the eyes and with a stretched exponential form $f(x) \sim
e^{-\alpha x^{\gamma}}$, where $(\gamma,\alpha)=(0.30,4.00)$ for the
lower 1/2 subset and (0.20,4.00) for the larger 1/2 subset
respectively. (b) The scaled conditional probability distribution
$P_{q} (\tau|\tau_{0})/\overline{\tau}$ vs $\tau/\overline{\tau}$
for the CMB stock with the threshold values $q$ ranging from 0.75 to
1.50 (0.750, 0.875, 1.000, 1.125, 1.250, 1.375 and 1.500) are
displayed with circles, squares, diamonds, triangle ups, triangle
lefts, triangle downs and triangle rights.  The closed symbols are
for the lower 1/2 subset, and  the open symbols are for the larger
1/2 subset. The dashed lines are for guiding the eyes and with a
stretched exponential form $f(x) \sim e^{-\alpha x^{\gamma}}$, where
$(\gamma,\alpha)=(0.30,4.00)$ for the lower 1/2 subset and
(0.20,3.50) for the larger 1/2 subset respectively.}
\end{figure}

\subsection{Mean Conditional Interval}
To further demonstrate the memory effect of the volatility return
intervals, we investigate the mean conditional return interval
$\langle\tau|\tau_{0}\rangle$, which is defined as the mean of the
volatility return intervals $\tau$ that immediately follow a given
$\tau_{0}$ subset. Fig. 5 shows the scaled mean conditional return
interval $\langle\tau|\tau_{0}\rangle/\overline{\tau}$ for the DTT
stock and the CMB stock with $q$=0.75, 1.00, 1.25. The closed
symbols are for the volatility return intervals and the open symbols
are for the shuffled data respectively. It is found that the
shuffled data $\langle\tau|\tau_{0}\rangle/\overline{\tau}$ almost
keeps a constant, i.e., there is no correlation in the shuffled
data. However, the clustering phenomena that small $\tau$ follows
small $\tau_{0}$ while large $\tau$ follows large $\tau_{0}$ are
observed for the volatility return intervals $\tau$. It further
demonstrates the clustering effect of the intervals for the Chinese
stock market. Therefore, it implies autocorrelation in  $\tau$. The
result is similar to that of the United States and the Japanese
stock markets \cite{yam05,wan06,wan07,jun08}.

\begin{figure}[htb]
\centering
\includegraphics[width=8.5cm]{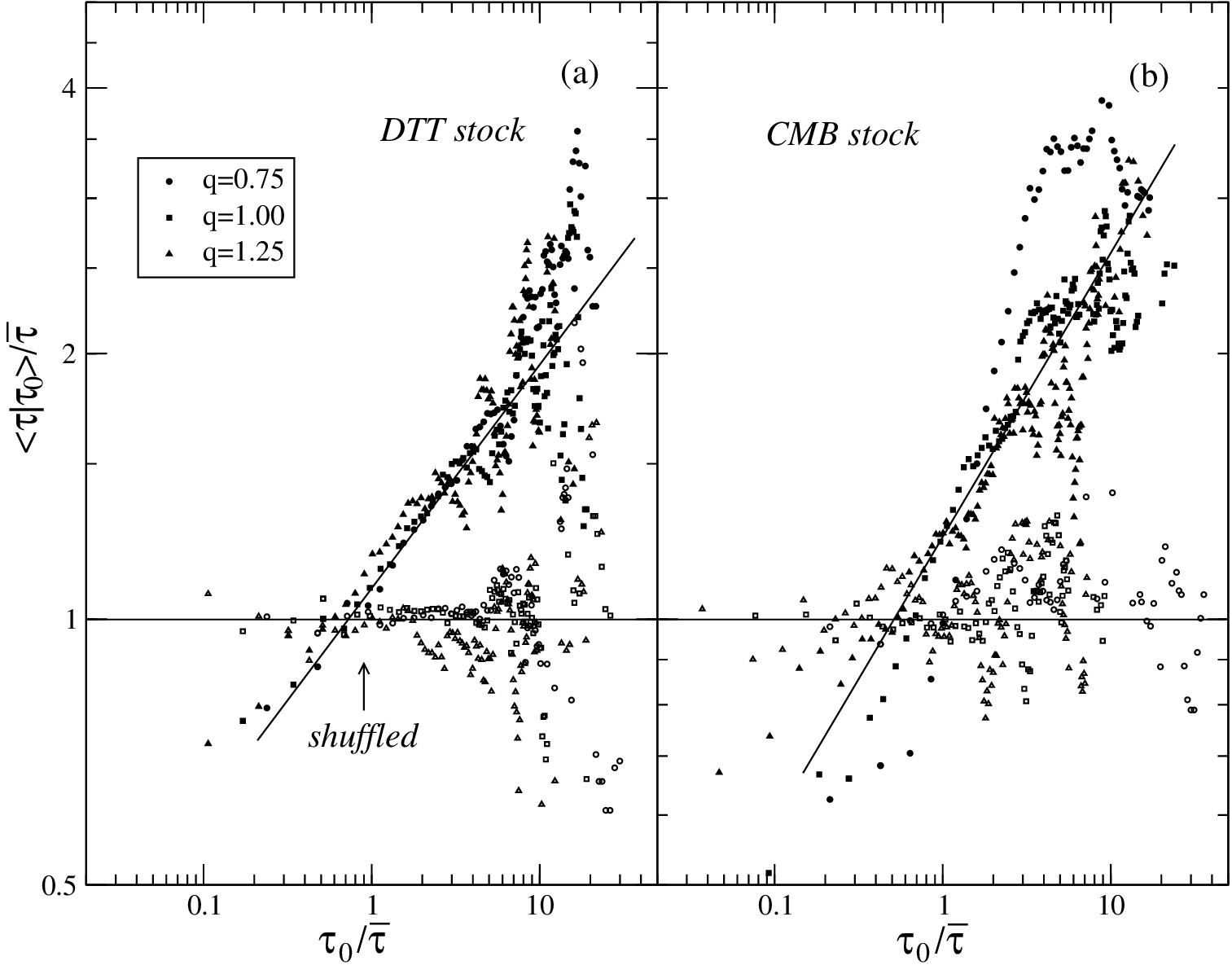}
\caption{\label{Fig:5} (a) The scaled mean conditional return
interval $\langle\tau|\tau_{0}\rangle/\overline{\tau}$ vs
$\tau_{0}/\overline{\tau}$ for the DTT stock with the threshold
values $q$=0.75, 1.00, 1.25 are displayed with circles, squares and
triangles. The closed symbols are for the volatility return
intervals $\tau$ and the open symbols are for the shuffled data. (b)
The scaled mean conditional return interval
$\langle\tau|\tau_{0}\rangle /\overline{\tau}$ vs
$\tau_{0}/\overline{\tau}$ for the CMB stock with the threshold
values $q$=0.75, 1.00, 1.25 are displayed with circles, squares and
triangles. The closed symbols are for the volatility return
intervals $\tau$ and the open symbols are for the shuffled data.}
\end{figure}

\subsection{Cluster Size Distribution of the Volatility Return Intervals}

To investigate the clustering phenomena in a more direct way, we
study the cumulative probability distribution of the cluster size of
$\tau$. The cluster size is obtained by calculating the successive
intervals with similar size \cite{wan06,wan07,jun08,vod08}. We
separate the data into two sets by the median data of $\tau$. The
data which are above (below) the median data is signed by $'+'$
($'-'$). Accordingly, $n$ consecutive $'+'$ or $'-'$ intervals form
a cluster and the corresponding cumulative probability distribution
is denoted as $p_{n+}(\tau)$ ($p_{n-}(\tau)$). Fig. 6 shows the
cumulative probability distribution $p_{n\pm}(\tau)$ for the DTT
stock and the CMB stock with $q$=0.75, 1.00 and 1.25. The open
symbols are for the $p_{n+}(\tau)$ and the closed symbols are for
the $p_{n-}(\tau)$. We find that the $p_{n-}(\tau)$ presents longer
tail persisting up to about $n=25$ than the $p_{n+}(\tau)$ does,
i.e., the relative small value of $\tau$ may form big clusters.
Similar clusters have been found in the United States and the
Japanese stock markets \cite{wan06,wan07,jun08,vod08}. It also
indicates long memory exists in $\tau$.

\begin{figure}[htb]
\centering
\includegraphics[width=8.5cm]{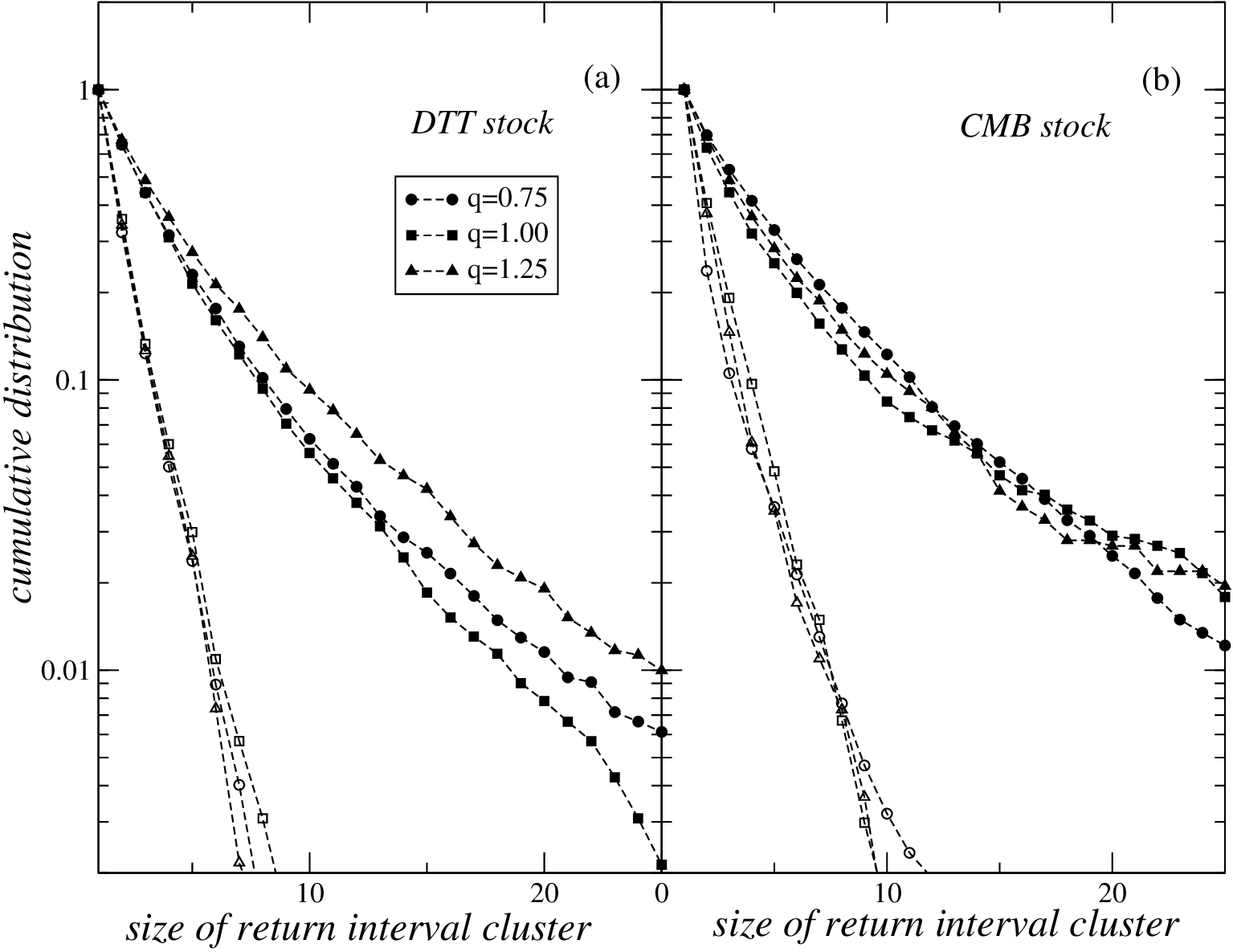}
\caption{\label{Fig:6} (a) Cumulative distribution of the cluster
size of $\tau$ for the DTT stock with the threshold values $q$=0.75,
1.00, 1.25 are displayed with circles, squares and triangles. The
open (closed) symbols are for the consecutive volatility return
intervals that are all above (below) the median interval records.
(b) Cumulative distribution of the cluster size of $\tau$ for the
CMB stock with the threshold values $q$=0.75, 1.00, 1.25 are
displayed with circles, squares and triangles. The open (closed)
symbols are for the consecutive volatility return intervals that are
all above (below) the median interval records.}
\end{figure}

\section{Persistence Probability}

To achieve deeper understanding of the memory effect of the
volatility return intervals, we investigate the persistence
probability, which has been systematically studied in nonequilibrium
dynamics such as phase ordering dynamics and critical dynamics \cite
{maj96,zhe98,ren03,ren04,ren05}. The idea of persistence is closely
related to the first passage time which has been widely studied in
physics, biology and engineering \cite
{ran00a,ran00b,ran00c,red01,zho05}. In general, the persistence
probability provides additional information to the autocorrelation.

The persistence probability $P_{+}(t)$ ($P_{-}(t)$) is defined as
the probability that $\tau(t'+\widetilde{t} )$ has always been above
(below) $\tau(t')$ in time $t$, i.e., $\tau(t'+\widetilde{t}
)>\tau(t')$ ($\tau(t'+\widetilde{t} )<\tau(t')$) for all
$\widetilde{t} < t$. The average is taken over $t'$. In Fig. 7, the
persistence probability distribution of $\tau$ is plotted in log-log
scale with $q$=0.75, 1.00 and 1.25. It is found that $P_{+}(t)$
decays much faster than $P_{-}(\tau)$ does. The $P_{-}(t)$ is
observed to decay by a power-law $P_{-}(t)\sim t^{-\beta}$. The
high-low asymmetry of the persistence probability is similar to that
of the volatility \cite {zhe98,ren03,ren04,ren05}. The persistence
exponents measured from the slopes of $P_{-}(t)$ with $q$=0.75, 1.00
and 1.25 are close and estimated to be $\beta=0.25 \pm 0.05$ for the
DTT stock, and $\beta=0.86 \pm 0.05$ for the CMB stock. The
exponents are far different from those of the random walk, where
both $P_{+}(t)$ and $P_{-}(t)$ show a power law behavior with a
persistence exponent $\beta = 0.50$. It further supports that
long-range correlation exists in $\tau$.

\begin{figure}[htb]
\centering
\includegraphics[width=8.5cm]{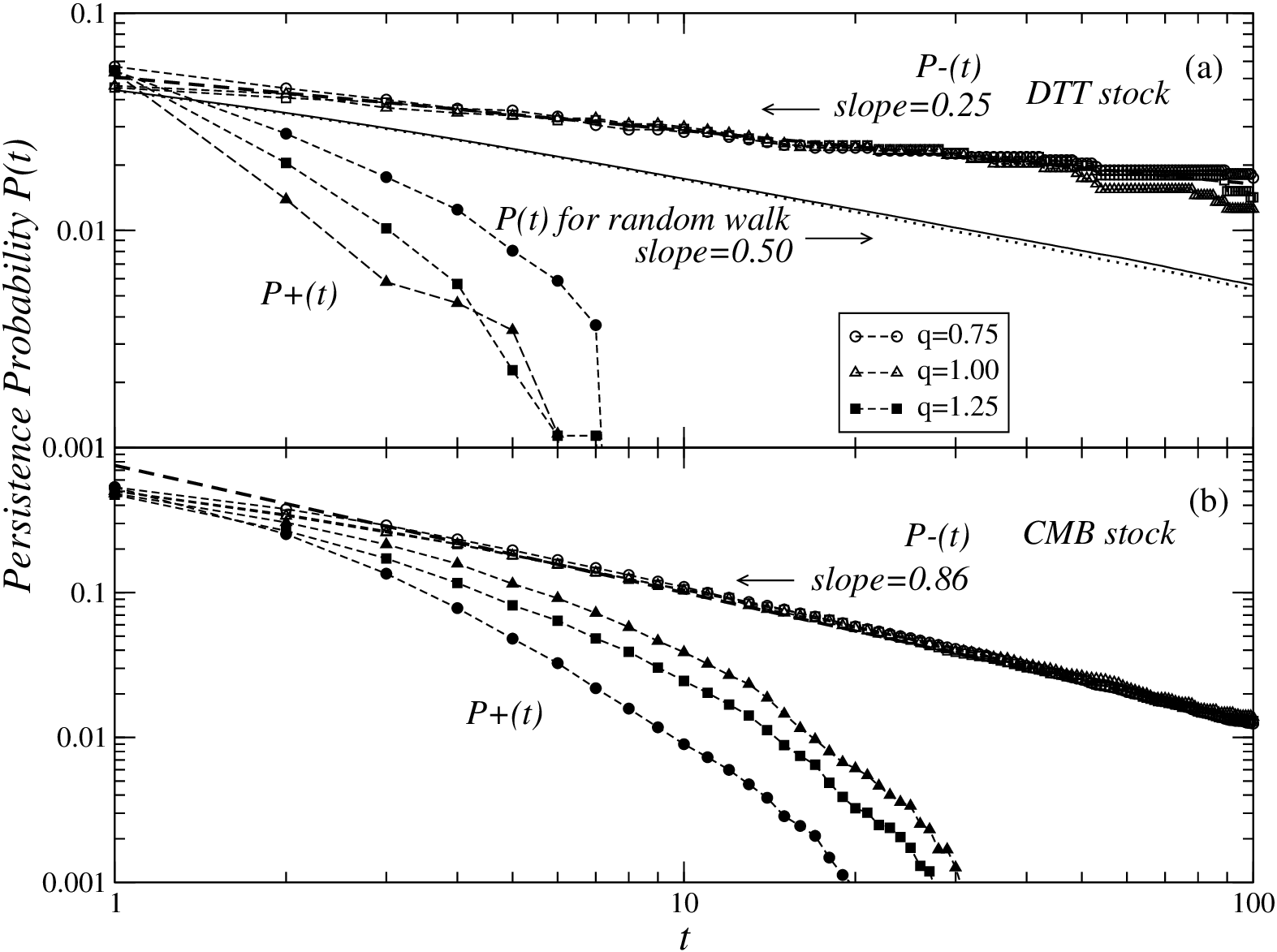}
\caption{\label{Fig:7} (a) Persistence probability of the volatility
return intervals $\tau$ for the DTT stock with the threshold values
$q$=0.75, 1.00, 1.25 are displayed with circles, squares and
triangles in log-log scale. The open (closed) symbols are for
$p_{-}(t)$ ($p_{+}(t)$). The solid line(the dotted line) is the
$p_{-}(t)$ ($p_{+}(t)$) for the random walk. (b) Persistence
probability of the volatility return interval $\tau$ for the CMB
stock with the threshold values $q$=0.75, 1.00, 1.25 are displayed
with circles, squares and triangles in log-log scale. The open
(closed) symbols are for $p_{-}(t)$ ($p_{+}(t)$).}
\end{figure}

\section{Conclusion}
In summary, we have investigated the probability distribution
function $P_{q}(\tau)$ of the volatility return intervals $\tau$ for
the Chinese stock market. Scaling behavior is observed after
$P_{q}(\tau)$ and $\tau$ are rescaled as
$P_{q}(\tau)=1/\overline{\tau} f(\tau/\overline{\tau})$. The scaling
curve can be fitted by a stretched exponential function $f(x) \sim
e^{-\alpha x^{\gamma}}$ with $\alpha = 0.39$ and $\gamma = 4.00$,
which is far different from the Poission distribution. It suggests
there exists memory in $\tau$. We then study the conditional
probability distribution $P_{q} (\tau|\tau_{0})$ and the mean
conditional return interval $\langle\tau|\tau_{0}\rangle$. The
results show that both the $P_{q} (\tau|\tau_{0})$ and the
$\langle\tau|\tau_{0}\rangle$ depend on the previous volatility
return intervals $\tau_{0}$. To obtain the clustering phenomena in a
more direct way, we investigate the cumulative probability
distribution of the cluster size of $\tau$. Clear clustering effect
is observed, especially for the relative small value of $\tau$. We
further investigate the persistence probability distribution of
$\tau$. It is found that $P_{-}(t)$ decays by a power law, with the
exponent $\beta=0.25 \pm 0.05$ for the DTT stock and $\beta=0.86 \pm
0.05$ for the CMB stock, which is far different from the value 0.5
for the random walk. The $P_{q} (\tau|\tau_{0})$, the
$\langle\tau|\tau_{0}\rangle$ and the cumulative probability
distribution of the cluster size of $\tau$ for the Chinese stock
market are similar to those obtained from the United States and the
Japanese stock markets. The persistence probability further confirms
the long memory of $\tau$ for the Chinese stock market. Compared
with the mature financial markets, we find that, as a emerging
market, the Chinese stock market may have some unique features,
however, it shares the similar scaling and long memory properties
for the volatility return intervals as the United States and the
Japanese stock markets.

\bigskip
{\textbf{Acknowledgments:}}

This work was supported by the National Natural Science Foundation
of China (Grant Nos. 10747138 and 10774080).

\bibliography{rint}
\end{document}